\documentclass[aps,prl,floats,floatfix,twocolumn]{revtex4}

\usepackage{ifthen}
\usepackage{ifpdf}
\usepackage{color}

\ifpdf
\usepackage{graphicx}
\usepackage{epstopdf}
\else
\usepackage{graphicx}
\usepackage{epsfig}
\fi

\usepackage{latexsym}
\usepackage{amsmath}
\usepackage{amssymb}
\usepackage{bm}
\usepackage{wasysym}

\newcommand{\mylabel}[1]{\label{#1}} 
\newcommand{\beq}{\begin{eqnarray}}
\newcommand{\eeq}{\end{eqnarray}} 
\newcommand{\be}[1]{\begin{eqnarray}\ifthenelse{#1=-1}
{\nonumber}{\ifthenelse{#1=0}{}{\mylabel{e#1}}}}
\newcommand{\ee}{\end{eqnarray}} 

\newcommand{\hide}[1]{}
\renewcommand{\cite}[1]{\textcolor{blue}{[\onlinecite{#1}}]} 

\newcommand{\ha}{\hat a}
\newcommand{\hn}{\hat n}
\newcommand{\hu}{\hat u}
\newcommand{\hv}{\hat v}
\newcommand{\hw}{\hat w}
\newcommand{\hS}{\hat S}

\begin{document}

\title{Dynamics of exciton-polaritons in a Josephson double dimer} 

\author{Christine Khripkov$^1$, Carlo Piermarocchi$^2$, and Amichay Vardi$^1$}

\affiliation{
\mbox{$^1$Department of Chemistry, Ben Gurion University of the Negev, Beer Sheva 84105, Israel}
\mbox{$^2$Department of Physics and Astronomy, Michigan State University, East Lansing 48824}
}

\begin{abstract}
We study the dynamics of exciton-polaritons in a double-well configuration. The system consists
of
two weakly coupled Bose-Josephson junctions, each corresponding to a different circular polarization of the polaritons, forming a {\it Josephson double dimer}. We show that the Josephson oscillation between the wells is strongly coupled to the polarization rotation and that consequently Josephson excitation is periodically exchanged between the two polarizations. Linearized analysis agrees well with numerical simulations using typical experimental parameters.
\end{abstract}

\maketitle

Recent experiments on  microcavity exciton-polaritons~\cite{savona1999}  have reported coherent oscillations and quantum self trapping characteristic of Bose Josephson Junctions, highlighting  the quantum fluid behavior of these quasiparticles~\cite{carusotto13}. Double quantum well systems were realized using coupled micropillars~\cite{abbarchi13,galbiati12} or by exploiting natural well width fluctuations that lead to a weak confinement for polaritons~\cite{lagoudakis10}. Several theoretical investigations anticipated these effects, and pointed out the richness of the dynamics that can be observed in these inherently out-of-equilibrium systems~\cite{wouters07,wouters08,sarchi08,shelykh08,solnyshkov09}. One  feature of microcavity polaritons is the presence of an  {\it intrinsic} Josephson effect~\cite{shelykh08}, resulting from the coupling of left and right circularly  polarized polaritons through an optical anisotropy term deriving from disorder or anisotropies in the microcavity structure~\cite{klopotowski06}. 

The simultaneous and coupled dynamics of left-right well and left-right polarization of polaritons naturally realizes a Josephson double dimer ~\cite{Strysz10,Strysz12a,Strysz12b,Chianca11}, which represents a minimal system to explore fundamental questions such as the emergence of thermodynamics in small quantum systems~\cite{Strysz10}. A polariton-based double dimer is particularly interesting because it allows for a fine control of the initial state preparation by a spatial and polarization shaping of the pulses, and it realizes a high temperature (compared to atoms), truly mesoscopic system. Moreover, ellipsometry measurements can provide a complete characterization of the polarization dynamics in the two wells. 

Polaritons possess  an inherent non-equilibrium nature due to their finite lifetime. We will show, however, that the radiative recombination of the polaritons does not affect the coherence of the two coupled oscillations, nor contribute to renormalization of the oscillation eigenfrequency. In addition, the internal spin structure of the polaritons  is  robust to spin decoherence processes since the non radiative dark exciton states corresponding to $J_z=\pm$ 2 are spectrally separated from the  polariton states with $J_z=\pm$1. 

\begin{figure}[t]
\includegraphics[width=0.8\hsize]{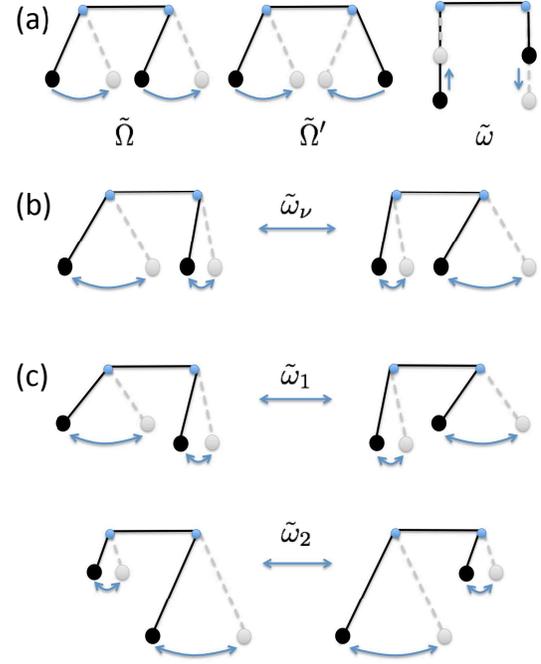}\\
\caption{Josephson oscillation modes of an exciton-polariton double-well condensate. Each pendulum represents a double-well Bose-Josephson junction with different circular polarization. From top: (a) Bogoliubov modes: antisymmetric (left), symmetric (center), and slow inter-polarization (right) oscillations; (b) Slow exchange of Josephson excitation (josons)
between the polarizations; (c) Natural adiabatic modes consisting of the combined exchange of particles and Josephson excitations between polarizations: opposite-phase particle- and joson oscillations with dominant energy transfer 
(top, $\tilde{\omega}_1$) 
and in-phase particle- and joson oscillations with dominant 
polarization transfer (bottom,  $\tilde{\omega}_2$).}
\label{f1}

\end{figure}


Signatures of  polarization precession were observed in Ref.~\cite{abbarchi13}.  However, a comprehensive picture of the coupling between interwell Josephson oscillations and  polariton polarization precession has not been obtained neither experimentally nor theoretically. In this work we apply the formalism of \cite{Strysz10,Strysz12a,Strysz12b}  and extend it to account for the finite polariton lifetime, to show that the intra-polarization Josephson oscillations (i.e. spatial oscillations of the same polarization species between the wells) and the inter-polarization Josephson oscillations (i.e. polarization-rotation within the wells) are in fact strongly coupled. Specifically, we study the case of strong intra-polarization coupling  and weak inter-polarization coupling.  Under such conditions, the two bosonic Josephson junctions corresponding to the two circular polarizations of the polaritons, exchange Josephson excitations ('josons')  as well as particles \cite{Strysz10}. Since the particle and joson oscillations are also coupled by the nonlinear frequency shifts, we find the natural frequencies of the collective particle-joson oscillation. The obtained frequencies agree well with numerical  simulations using characteristic experimental parameters.

{\bf Bose-Hubbard model and pseudospin representation.-- } We model the polariton double-well system using the Bose-Hubbard tight-binding four-mode model of two weakly coupled dimers,
\begin{eqnarray}
\label{Hamiltonian}
{\hat H}&=&\left(-\frac{\Omega}{2}\sum_{\sigma} \ha_{\sigma,L}^\dag \ha_{\sigma,R}
%
-\frac{\omega}{2}\sum_{\alpha} \ha_{+,\alpha}^\dag \ha_{-,\alpha}+ H.c.\right)\nonumber\\
~&~&+U\sum_{\sigma,\alpha} \hn_{\alpha,\sigma}(\hn_{\alpha,\sigma}-1)~,
\end{eqnarray}
where $\ha_{\sigma,\alpha}$ are the annihilation operators for quasiparticles with circular polarization $\sigma=\pm$  in well, $\alpha=\{L,R\}$, $\Omega$ and $\omega$ are respectively the intra-polarization and inter-polarization coupling frequencies, and $U$ corresponds to the exciton-polariton interaction strength. We assume that $\Omega\gg\omega$ in agreement with current experimental conditions \cite{abbarchi13}. 

We map the Hamiltonian (\ref{Hamiltonian}) into a spin problem \cite{Tikhonenkov07} by ordering $(\ha_{+,L},\ha_{+,R},\ha_{-,R},\ha_{-,L})\leftrightarrow(\ha_1,\ha_2,\ha_3,\ha_4)$ and defining fifteen SU(4) generators $\hu_{j,k}\equiv\ha^\dag_k\ha_j+\ha_j^\dag\ha_k$,   $\hv_{j,k}\equiv(\ha^\dag_k\ha_j-\ha_j^\dag\ha_k)/i$ and $\hw_l=\sqrt{2/[l(l+1)]}\left(\sum_{j=1}^l \hn_j - l\hn_{l+1}\right)$ with $1\leq k<j\leq 4$, $1\leq l \leq 3$, and $\hn_j\equiv\ha_j^\dag\ha_j$. Constructing a pseudo-spin vector $\hat{\bf S}=(\hu_{2,1}, \hu_{3,2},\hu_{4,3},\hu_{3,1},\hu_{4,2},\hu_{4,1},\hv_{2,1},\cdots,\hv_{4,1},\hw_1,\hw_2,\hw_3)$ from these generators (in analogy to ${\bf S}=(u,v,w)$ for a two-level system), we obtain,
\begin{equation}
{\hat H}=-\frac{\Omega}{2}(\hS_1+\hS_3)-\frac{\omega}{2}(\hS_2+\hS_6)+\frac{U}{2} (\hS_{13}^2+\hS_{14}^2+\hS_{15}^2)~,
\label{HamSpin}
\end{equation}
where we have eliminated a $UN(N/4-1)$ $c$-number term which does not affect the dynamics. The finite lifetime of the polaritons $\tau$ is accounted for by solving the master equation for the $N$-polaritons density matrix $\hat\rho$,
\begin{equation}
i\frac{d\hat\rho}{dt}=\left[{\hat H},{\hat\rho}\right]+i{\cal L}{\hat\rho}~,
\label{master}
\end{equation}
with the Lindblad operator,
\begin{equation}
{\cal L}=-\frac{\gamma}{2}\sum_{j=1}^4 \left( \ha_j^\dag\ha_j{\hat\rho}+{\hat\rho}\ha_j^\dag\ha_j-2\ha_j{\hat \rho}\ha_j^\dag \right)
\label{loss}
\end{equation}
where $\gamma=1/\tau$. The dynamical equations for the $S_i$ operators thus read,
\begin{eqnarray}
i{\dot \hS}_i&=&-\frac{\Omega}{2}\sum_{j=1,3}\sum_{k=1}^{15} c_{ij}^k \hS_k -\frac{\omega}{2}\sum_{j=2,6}\sum_
{k=1}^{15}
 c_{ij}^k \hS_k \nonumber\\
~&~&+ U\sum_{j=13}^{15}\sum_
{k=1}^{15} 
c_{ij}^k \hS_j \hS_k- i\gamma \hS_i~,
\end{eqnarray} 
where $c_{ij}^k$ are SU(4) structure constants.

The reduced SU(2) representation for the two Bose-Hubbard dimers with circular polarization $\sigma=\pm$ are extracted from the SU(4) pseudospin as ${\hat{\bf S}}_+ = (\hS_1,\hS_7,\hS_{13})$ and ${\hat{\bf S}}_- = (\hS_3,-\hS_9,\sqrt{1/3}\hS_{14}-\sqrt{2/3}\hS_{15})$ whereas the corresponding particle number in each polarization is 
${\hat N}_{+(-)}={\hat n}_{1(4)}+{\hat n}_{2(3)}$. 
We thus obtain two Bloch/Poincar\'e vectors ${\bf s}_\sigma=\langle\hat{\bf S}_\sigma\rangle/ |\langle\hat{\bf S}_\sigma\rangle|$ where $s_{\sigma,z}$ corresponds to the normalized spatial population imbalance within the polarization $\sigma$, and $s_{\sigma,x},s_{\sigma,y}$ correspond respectively, to the real and imaginary parts of the pertinent spatial coherence.

{\bf Mechanical equivalent and Josephson oscillation modes.-- } The low energy spectrum for each of the two weakly-coupled Bose-Hubbard dimers with different polarizations, consists of harmonic-oscillator levels with Josephson frequency spacing ${\tilde\Omega}_\sigma=\sqrt{\Omega(\Omega+2UN_\sigma)}$ \cite{Chuchem10}. The four-mode system in this linear regime, is thus equivalent to coupled non-rigid pendulums which can exchange their length, corresponding to the number of particles in each dimer $N_\sigma\equiv\langle {\hat N}_\sigma\rangle$, as well as their oscillation amplitude, corresponding to the excitation of each oscillator. The degree of this Josephson-excitation is quantified by the respective {\em number of josons},
\begin{equation}
\nu_\sigma = \frac{E_\sigma - E_{\sigma,g}}{\tilde\Omega_\sigma}= \frac{1}{\tilde\Omega_\sigma} \left[-\frac{\Omega}{2} \langle \hS_{\sigma,x}\rangle+\frac{U}{2} \langle\hS_{\sigma,z}^2\rangle+\frac{\Omega N_\sigma}{2}\right]
\end{equation}
where $E_\sigma$ and $E_{\sigma,g}=-\Omega N_\sigma /2$ denote respectively the energy and ground-state energy of dimer $\sigma$. While the total number of josons $\nu=\nu_+ + \nu_-$ is not strictly conserved, conservation of energy implies it is an adiabatic invariant in the linear regime.

The Bogoliubov linearized collective-oscillation modes for the Hamiltonian (\ref{HamSpin}) are illustrated in Fig~\ref{f1}a. In agreement with the coupled-pendulum analogy, they include: (i) An antisymmetric mode at the single-pendulum frequency ${\tilde\Omega}=\sqrt{\Omega(\Omega+UN)}$ where the spatial oscillations  of the two polarizations are in-phase; (ii) A symmetric mode with spatial oscillations of opposite phase in the two polarizations, at frequency ${\tilde\Omega}'=\sqrt{(\Omega+\omega)(\Omega+\omega+UN)}$; and (iii) Slow particle-oscillations between the two polarizations, maintaining a fixed total population $N$, with  frequency 
\begin{equation}
{\tilde\omega}=\sqrt{\omega(\omega+UN)}~.
\end{equation}

The beating of the two fast Bogoliubov modes at the difference frequency,
$\omega_\nu={\tilde\Omega}'-\tilde{\Omega}\approx(\omega/{\tilde\Omega})(\Omega+UN/2)$
corresponds to the slow exchange of josons between the two polarizations (Fig.~\ref{f1}b). As shown in Refs.~\cite{Strysz10,Strysz12a,Strysz12b}  using a series of Holstein-Primakoff \cite{HPT} and Bogoliubov transformations, this beat mode serves as an independent adiabatic oscillation mode whose frequency is modified as 
\begin{equation}
{\tilde\omega}_\nu=\sqrt{\omega_\nu(\omega_\nu+U_\nu \nu)}~,
\end{equation}
by the effective attractive interaction between josons $U_\nu=-(U/4)(4\Omega+UN)/(\Omega+UN)$.

The proper description of the exciton-polariton bosonic Josephson system is thus in terms of two Bose-Hubbard dimers of different polarizations {\em which can exchange  both particles and excitations}. This is a far more intricate picture than the prevalent view of independent, non-coupled modes for spatial/extrinsic oscillations and polarization/intrinsic oscillations.
Furthermore, the particle-oscillations (at frequency $\tilde\omega$) and excitation-oscillations (at frequency $\tilde\omega_\nu$) are also coupled because the frequency of oscillation within each polarization depends on the number of polaritons in it.  Using the mechanical analogy, the pendulum frequencies change as they exchange their length. The effective particle-joson coupling is $U_c=(U/2)(\Omega/{\tilde\Omega})\sqrt{N\nu}$. Consequently, the final oscillation modes involve both particle and joson exchange. Diagonalization  of the effective Hamiltonian for coupled particle- and joson oscillators eventually yields the natural frequencies \cite{Strysz10,Strysz12a,Strysz12b},
\begin{equation}
\tilde{\omega}_{1,2}^2=\frac{{\tilde\omega}^2+{\tilde\omega}_\nu^2}{2}\mp\left [ \left( \frac{{\tilde\omega}^2-{\tilde\omega}_\nu^2}{2}\right)^2+\frac{\omega\omega_
\nu\Omega U^2 N\nu}{\Omega+UN}\right]^{1/2}~.
\label{naturalfrequencies}
\end{equation}

The two natural slow modes are shown in Fig.~\ref{f1}c.  The slower oscillation at frequency ${\tilde\omega}_1$, constitutes predominantly energy-exchange (i.e. joson oscillations) accompanied by small population oscillations. The particle- and joson oscillations in this mode are of opposite phase, i.e. the joson imbalance $\Delta\nu\equiv\nu_+ - \nu_-$ is maximized when the particle imbalance $\Delta N\equiv N_+ - N_-$ is at minimum. By contrast the faster mode at frequency  ${\tilde\omega}_2$, has dominant particle oscillations accompanied by minor joson oscillations. In this case, particles and josons oscillate in-phase.

{\bf Effect of losses.-- }
The loss mechanism described in Eq.~(\ref{master}) with the Lindbladian (\ref{loss}) and a polariton lifetime of 33 ps was shown to accurately describe the experimental observations for coupled micropillars \cite{abbarchi13}. Dephasing times for exciton-polariton systems are of order of hundreds of ps and are consequently neglected. Since the loss Lindbladian equally dampens populations and coherences, it does not produce any frequency shifts other than the change in the chemical potential incurred by the decay of $N(t)$ and $\nu(t)$. The natural frequencies of Eq.~(\ref{naturalfrequencies}) thus slowly vary in its presence as,
\begin{equation}
\tilde{\omega}_{1,2} \rightarrow \tilde{\omega}_{1,2}(N(t),\nu(t))-i\gamma~,
\end{equation}
with $N(t)=N(0)e^{-t/\tau}$ and $\nu(t)=\nu(0)e^{-t/\tau}$.

\begin{figure}[t]
\includegraphics[width=0.8\hsize]{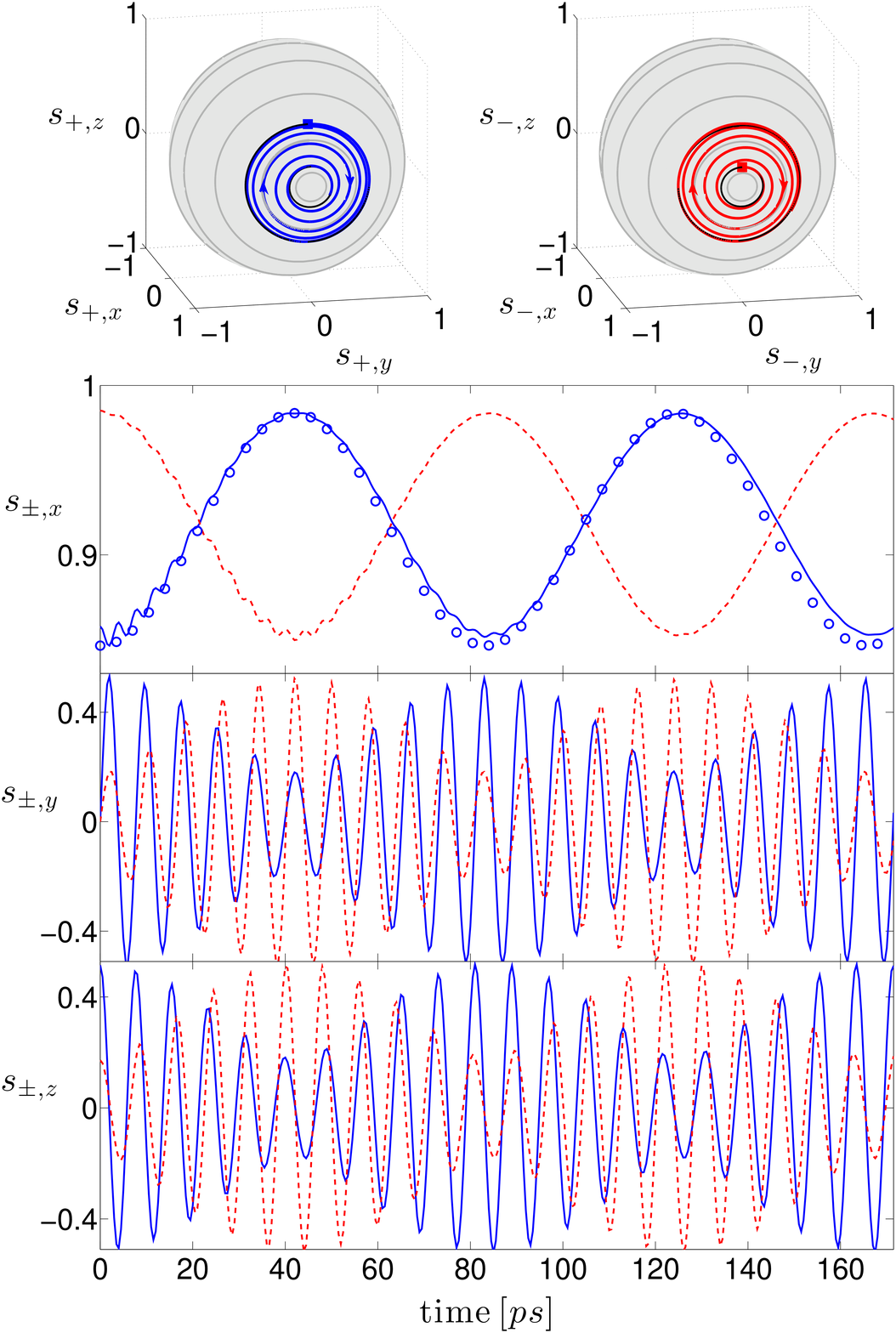}\\
\caption{Exchange of Josephson excitations between polarizations: (top) Mean-field evolution of the reduced Bloch vectors  ${\bf s}_+$ (left) and ${\bf s}_-$ (right). Symbols mark the initial state; (bottom) Components of ${\bf s}_+$ (solid blue line) and ${\bf s}_-$ (dashed red line). Circles mark oscillation at the loss-corrected frequency $\tilde{\omega}_1(t)$. Parameters are $U=0.1\mu$eV, $\Omega=500\mu$eV, $\omega=50\mu$eV, $\tau=50$ps, $N=500$, $\nu=20$. Initial conditions correspond to $\Delta N= -13.6$ and $\Delta\nu=16$ with zero-relative phase between polarizations and between the spatial components of each polarization, designed to excite the $\tilde{\omega}_1$ collective oscillation. The number of particles at the end of the simulation is $N\sim 16$.}
\label{f2}
\end{figure}


\begin{figure}[t]
\includegraphics[width=0.7\hsize]{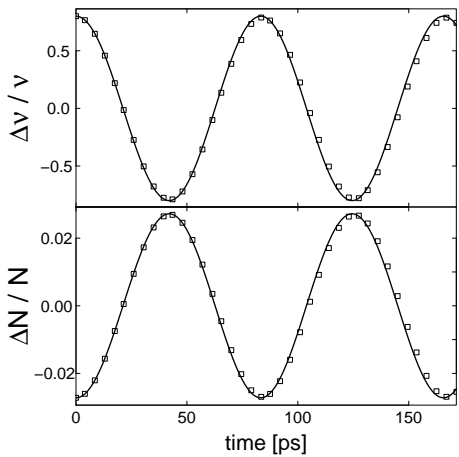}\\
\caption{Dynamics of the normalized joson imbalance $\Delta\nu/\nu$ and particle imbalance $\Delta N /N$ between the two polarizations, for the parameters and preparation of Fig~\ref{f2}. Squares denote oscillations at the loss-corrected frequency $\tilde{\omega}_1$.}
\label{f3}
\end{figure}


{\bf Numerical results.-- }
In order to assess the experimental feasibility of observing composite particle-joson oscillations between the polarizations of an exciton-polariton condensate in a double-well potential, we carry out numerical simulations with characteristic experimental parameters. The tunnel coupling between the confined polaritonic ground states is taken to be $\Omega=500\mu{\rm eV}$ and the polarization splitting is $\omega=50\mu{\rm eV}$. The interaction constant is $U=0.1\mu{\rm eV}$ and the number of polaritons is $N=500$.  The characteristic interaction parameters are thus $u_\Omega=2UN/\Omega = 0.2$ and $u_\omega=2UN/\omega = 2$. The polariton lifetime is estimated as  $\tau=50$ ps. 

For the linear regime studied in this work, a classical mean-field solution in which the operators $\hat{\bf S}$ are replaced by $c$-numbers ${\bf S}$, provides an accurate description of the quantum system. We have verified this approximation by conducting full quantum numerical simulations at the same values of the interaction parameters $u_{\Omega,\omega}$, obtaining excellent agreement for the duration of the simulation with particle number $N\sim 50$. Since quantum fluctuations diminish as $1/N$, the quantum-classical agreement would even improve for $N=500$ (where quantum calculations are precluded due to computational limitations).

In Fig.~\ref{f2} we plot the time evolution of the reduced Bloch vectors $\bf s_{\pm}$ with a preparation designed to excite the $\tilde{\omega}_1$ collective mode: Both the relative-phase between the two polarizations and the internal relative-phases of the two dimers are set to zero, with an initial population (particle) and energy (joson) imbalance of opposite signs. As expected, we observe an exchange of energy between the two polarizations, wherein the amplitude of the internal Josephson oscillations slowly alternates between the spheres at the expected frequency 
$\tilde{\omega}_1$. The resulting inter-polarization population- and energy-oscillations are shown in Fig.~\ref{f3} with the $\pi$ phase-difference between them and the larger joson oscillation amplitude  expected for this composite natural mode.

\begin{figure}[t]
\includegraphics[width=0.8\hsize]{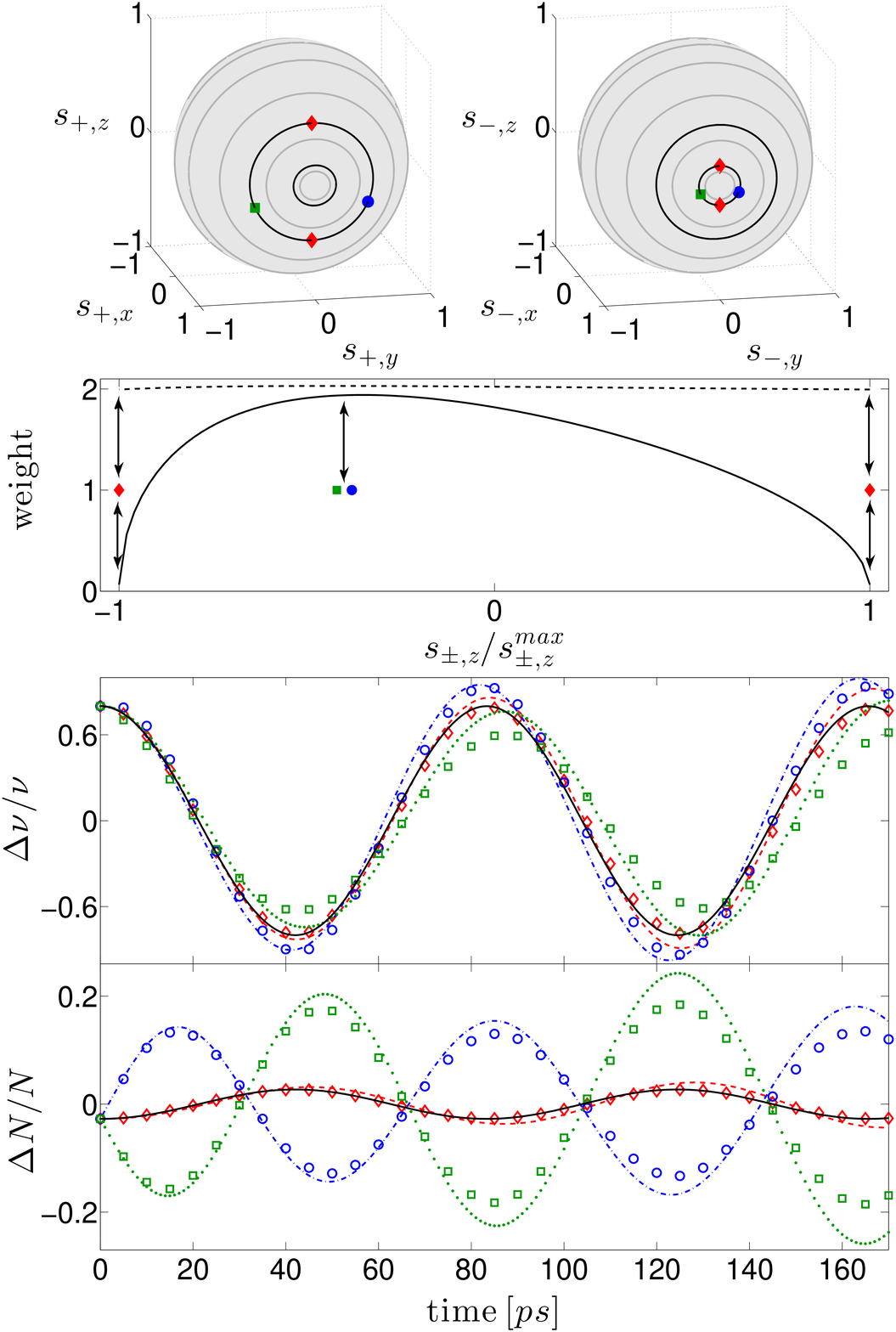}\\
\caption{Comparison of different preparations with the same initial $\Delta N, \Delta\nu$ as in Fig.~\ref{f3}: (top) The initial reduced Bloch vectors ${\bf s}_\pm(t=0)$ ; (middle) Weights
(not normalized)
of the $\tilde{\omega}_1$ mode (dashed) and $\tilde{\omega}_2$ mode (solid) as a function of the relative intrapolarization particle imbalance $s_{\pm,z}/s_{\pm,z}^{max}$ where $s_{\pm,z}^{max}$ is the maximum value of $s_{\pm,z}$ within the fixed $\Delta N, \Delta\nu$ subspace; (bottom) Dynamics of particle- and joson-imbalance. 
 Parameters are the same as in Fig.~\ref{f2}. Symbols are numerical mean-field results whereas lines depict the pertinent combination of slow natural mode oscillations at $\tilde{\omega}_{1,2}$, with the weights from the middle panel.}
\label{f4}
\end{figure}


We compare the joson-particle oscillations observed here with Fig. 2 of Ref.~\cite{shelykh08}. While the results may superficially appear similar, the spatial beating in \cite{shelykh08} emerges simply from two {\em uncoupled}  ($\omega=0$) polarization dimers oscillating independently at different frequencies due to the different shifts caused by their unequal fixed population. Such dynamics involves neither particle exchange nor energy exchange between the polarizations. Detecting one polarization, an experimentalist  would simply observe standard two-mode oscillations. By contrast, here we have true four-mode dynamics with both population and energy flowing back an forth from one polarization to the other. Moreover, this flow is non-trivial with the more populated polarization carrying less excitation for the $\tilde{\omega}_1$ oscillation mode depicted in Fig.~\ref{f2} and Fig.~\ref{f3} but with opposite behavior for the $\tilde{\omega}_2$
mode.

In Fig.~\ref{f4} we compare the mean-field evolution of various preparations which have the same initial interpolarization population- and energy-difference. While the preparation in which all four mode amplitudes have the same phase excites purely the $\tilde{\omega}_1$ oscillation, other preparations excite also the $\tilde{\omega}_2$ mode.  The degree of this mode-mixing is determined purely by the internal population imbalance within the polarizations $s_{\pm,z}$. Good agreement is obtained with the adiabatic natural mode analysis \cite{Strysz10,Strysz12a,Strysz12b} which we have modified to include losses.

{\bf Conclusion.-- } The dynamics of exciton-polariton Josephson junctions is more complex than that of independent intrapolarization and interpolarization Josephson oscillators. Indeed the natural slow modes for interpolarization oscillations involve both particle oscillation and energy oscillation through the exchange of intrapolarization Josephson excitation. Furthermore, these slow modes are coupled, resulting in natural frequencies which involve both particle and excitation exchange. Numerical simulations with typical parameters, including radiative-recombination losses, indicate that such behavior can be observed within current experimental measurement times.

{\bf Acknowledgments.-- }
We gratefully acknowledge valuable discussions with James Anglin, Doron Cohen, and Martin Strysz. This research was supported by the the United States-Israel Binational Science Foundation (BSF grant No. 2008141) and the Israel Science Foundation (ISF grant No. 346/11).



\begin{thebibliography}{99}

\bibitem{savona1999}
V. Savona, C. Piermarocchi, A.  Quattropani, P.  Schwendimann,  and F. Tassone, Phase Transit. {\bf 68}, 169  (1999).

\bibitem{carusotto13}
I. Carusotto and C. Ciuti, Rev. Mod. Phys. {\bf 85}, 299 (2013).

\bibitem{abbarchi13}
M. Abbarchi {\it et al.}, Nature Phys. {\bf 9}, 275 (2013).  

\bibitem{galbiati12}
M.  Galbiati {\it et al.}, Phys. Rev. Lett. {\bf 108}, 126403 (2012)

\bibitem{lagoudakis10}
K. G. Lagoudakis, B. Pietka, M. Wouters, R. Andr\'e, and B. Deveaud-Pl\'edran, Phys. Rev. Lett. {\bf 105}, 120403 (2010).

\bibitem{wouters07}
M. Wouters and I. Carusotto, Phys. Rev. Lett. {\bf 99}, 140402 (2007).

\bibitem{wouters08}
M. Wouters, Phys. Rev. B {\bf 77}, 121302 (2008).


\bibitem{sarchi08}
D. Sarchi, I. Carusotto, M. Wouters, and V. Savona, Phys. Rev. B {\bf 77}, 125324 (2008).

\bibitem{shelykh08}
I. A. Shelykh, D. D. Solnyshkov, G. Pavlovic, and G. Malpuech, Phys. Rev. B {\bf 78}, 041302 (2008).


\bibitem{solnyshkov09}
D. D. Solnyshkov, R. Johne, I. A. Shelykh, and G. Malpuech, Phys. Rev. B {\bf 80}, 235303 (2009).

\bibitem{klopotowski06}
{\L}. K{\l}opotowski {\it et al.}, Solid State Commun. {\bf 139}, 511 (2006).


\bibitem{Strysz10}
M. P. Strysz and J. R. Anglin,
Phys. Rev. A {\bf 81}, 043616 (2010).


\bibitem{Strysz12a}
M. P. Strysz and J. R. Anglin,
Phys. Rev. A  {\bf 85}, 053610 (2012).

\bibitem{Strysz12b}
M. P. Strysz and J. R. Anglin,
Phys. Scr. {\bf T151}, 014046 (2012).

\bibitem{Chianca11}
C. V. Chianca and M. K. Olsen,
Phys. Rev. A {\bf 83}, 043607 (2011).

\bibitem{Tikhonenkov07}
I. Tikhonenkov, J. R. Anglin, and A. Vardi,
Phys. Rev. A {\bf 75}, 013613 (2007).

\bibitem{Chuchem10}
M. Chuchem, K. Smith-Mannschott, M. Hiller, T. Kottos, A. Vardi, and D. Cohen,
Phys. Rev. A {\bf 82}, 053617 (2010).

\bibitem{HPT}
T. Holstein and H. Primakoff, Phys. Rev. {\bf 58}, 1098 (1949).

\end{thebibliography}
\end{document}